\documentclass[journal]{IEEEtran}

\usepackage{amssymb, latexsym, amsmath, graphicx, epstopdf, here, xcolor, cite, enumitem}

\usepackage{kbordermatrix}
\renewcommand{\kbldelim}{[}
\renewcommand{\kbrdelim}{]}

\usepackage{tikz}
\usetikzlibrary{
	patterns,
	calc,
	fit,
	arrows,
	plotmarks,
	shadows,
	chains,
	shapes
}

\newtheorem{thm}{Theorem}
\newtheorem{cor}{Corollary}
\newtheorem{lem}{Lemma}
\newtheorem{prop}{Proposition}
\newtheorem{defi}{Definition}

\newcommand{\Rin}{\mathcal{R}_{\sf i}}
\newcommand{\Rout}{\mathcal{R}_{\sf o}}
\newcommand{\Routt}{\mathcal{R}_{\sf o}^{\sf OWC}}
\newcommand{\Cowc}{\mathcal{C}_{\sf OWC}}

\IEEEoverridecommandlockouts

\title{Classes of Full-Duplex Channels With Capacity Achieved Without Adaptation}

\author{
	Daewon Seo, Anas Chaaban, Lav R.~Varshney, and Mohamed-Slim Alouini
	\thanks{D.~Seo is with the Department of Electrical and Computer Engineering, University of Wisconsin-Madison, Madison, WI 53706 USA (e-mail: dseo24@wisc.edu). A.~Chaaban is with the School of Engineering, University of British Columbia, Kelowna, BC V1Y 1V7 Canada (e-mail: anas.chaaban@ubc.ca). L.~R.~Varshney is with the Coordinated Science Laboratory and the Department of Electrical and Computer Engineering, University of Illinois at Urbana-Champaign, Urbana, IL 61801 USA (e-mail: varshney@illinois.edu). M.-S.~Alouini is with the Division of Computer, Electrical, and Mathematical Sciences and Engineering, King Abdullah University of Science and Technology, Thuwal, Saudi Arabia (e-mail: slim.alouini@kaust.edu.sa).}
	\thanks{This work was completed while Seo was at University of Illinois at Urbana-Champaign. Parts of this work were presented at the 2013 IEEE International Symposium on Information Theory \cite{Varshney2013c} and the 2017 IEEE International Symposium on Information Theory \cite{ChaabanVA2017}.}
}

\begin{document}

\maketitle

\begin{abstract}
Full-duplex communication allows a terminal to transmit and receive signals simultaneously, and hence, it is helpful in general to adapt transmissions to received signals. However, this often requires unaffordable complexity. This work focuses on simple non-adaptive transmission, and provides two classes of channels for which Shannon's information capacity regions are achieved without adaptation. The first is the injective semi-deterministic two-way channel that includes additive channels with various types of noises modeling wireless, coaxial cable, and other settings. The other is the Poisson two-way channel, for which we show that non-adaptive transmission is asymptotically optimal in the high dark current regime.
\end{abstract}


\section{Introduction} \label{sec:intro}
Full-duplex communication is gaining popularity for next-generation networks since it has potential to double the spectral efficiency. Recent applications, such as cloud services, real-time gaming, live streaming video, and augmented reality, require high-speed and symmetric performance of uplink and downlink.  This need has recently led to accelerated research on full-duplex systems. Practical examples of full-duplex include wireless settings \cite{ChoiJSLK2010, JainCKBSSLKS2011, Duarte2012, SabharwalSGBRW2014}, as well as wireline settings, e.g.~over coaxial cable \cite{CoomansCM2018, BerscheidH2019}.

Therefore there is a growing interest in studying fundamental information-theoretic limits of the full-duplex model, where a terminal is allowed to transmit and receive simultaneously. It was first studied by Shannon \cite{Shannon1961} under the name of \emph{two-way channel} (TWC), and inner and outer bounds for the capacity region of general discrete alphabet memoryless TWCs were given.

The inner bound is achieved using non-adaptive coding, i.e., the transmit signal at each terminal is determined only by the message, irrespective of the received signals up to that time instant. Therefore, the signals to be sent at terminals are independent of each other. The outer bound, on the other hand, allows those signals to be dependent. Several works have focused on tightening those bounds \cite{HekstraW1989, Schalkwijk1982} (cf.~\cite[Chap.~17]{ElGamalK2011}). The fact that Shannon's bounds do not coincide in general \cite{Dueck1979} implies an important technique for coding, namely \emph{adaptation}. However, if they do coincide for a given channel, this implies that adaptive encoding is not needed to achieve capacity for this particular channel.

Since communication strategies without adaptation simplify engineering system design, there is growing interest in determining classes of channels for which the capacity region can be achieved without adaptation \cite{ChengD2014}. These channels include modulo-$2$ adder channel, the class of symmetric discrete alphabet memoryless TWCs \cite{Shannon1961}, and the additive white Gaussian noise (AWGN) TWC \cite{Han1984}. It is also known that for the AWGN TWC with additive interference and transmitter side information, the capacity is achieved by dirty paper coding, hence, without adaptation \cite{Khosravi-FarsaniR2011}. It has recently been established that adaptation is not needed to achieve capacity for a broad class of channels with certain symmetric properties \cite{WengSAL2019, WengAL2019, SabagP2018}. A more detailed survey is provided in \cite[Chap.~2]{ChaabanS2015}.

Here, we not only study some discrete-time TWCs similar to above, but also the Poisson channel, which is continuous-time and has been used to model optical communication \cite{GhassemlooyPR2018} as well as communication over a bacterial cable \cite{MichelusiM2015}. The one-way capacity of the continuous-time Poisson channel has been studied with \cite{ShamaiL1993} and without \cite{Davis1980, Wyner1988a, Wyner1988b} bandwidth constraints on input waveforms. Beyond the point-to-point channel, Poisson multiple-access \cite{LapidothS1998}, broadcast \cite{LapidothTU2003}, and interference \cite{LaiLS2015} channels have also been studied. As far as we know, the Poisson TWC is unstudied.

This paper focuses on TWCs for which capacity regions are achievable without adaptation, and therefore system design remains simple without losing information rate. We first consider a class of injective semi-deterministic (ISD) TWCs in Sec.~\ref{sec:ISD}. It captures many important two-way communication models, for example, additive models with various types of independent noises, such as Gaussian, generalized exponential, and Cauchy noises, and even some input-dependent noise, as given in Sec.~\ref{sec:ISD_example}. Those models arise in wireless and coaxial cable full-duplex communication. In addition, we study the Poisson TWC in Sec.~\ref{sec:Poisson}, which is a model for optical communication. It is shown that the capacity region is asymptotically rectangular as dark current intensity goes to infinity, which implies that adaptation is asymptotically useless.

Since the first presentation of some of our results \cite{Varshney2013c, ChaabanVA2017} about ISD TWCs, more general characterizations of two-way channel capacity regions have been established \cite{WengSAL2019}. Notwithstanding, the purpose of this paper is to show that explicit capacity formulas for such ISD channel family are available and such non-adaptive transmission is insightful for the design of full-duplex communication schemes over realistic channel models arising from important communication systems. Another contribution concerns the asymptotic capacity region of the Poisson TWC, a member of continuous-time channels for which obtaining capacity formulas is largely unstudied. Our contributions are summarized as follows.
\begin{itemize}
\item We provide a sufficient condition under which the capacity region of general TWC is established. A class of ISD TWCs, which arises in many practical full-duplex systems, is shown to satisfy the condition and yield explicit capacity formulas without adaptive transmission.

\item We give examples of multiplicative channels and additive channels with input independent/dependent noise whose capacity region is achieved without adaptation, and we express their capacity region.

\item We define the capacity region of the Poisson TWC, which arises in optical full-duplex communication, and prove that it is asymptotically achievable in the large dark current regime without adaptation. Moreover, a simple asymptotic capacity expression is derived.
\end{itemize}

\section{System Model and Preliminaries} \label{sec:model}
\subsection{Channel Model} \label{subsec:channel_model}
\begin{figure}
	\centering
	\begin{tikzpicture}
	\node (1) at (0,0) [rectangle,draw,minimum height=1.5cm] {1};
	\node (x1) at ($(1)+(1,.5)$) {$X_1$};
	\node (y1) at ($(1)+(1,-.5)$) {$Y_1$};
	\draw[->] ($(1.east)+(0,.5)$) to (x1);
	\draw[<-] ($(1.east)+(0,-.5)$) to (y1);
	
	\node (2) at (6,0) [rectangle,draw,minimum height=1.5cm] {2};
	\node (x2) at ($(2)+(-1,-.5)$) {$X_2$};
	\node (y2) at ($(2)+(-1,.5)$) {$Y_2$};
	\draw[->] ($(2.west)+(0,-.5)$) to (x2);
	\draw[<-] ($(2.west)+(0,.5)$) to (y2);
	
	\node (f1) at ($(1)+(3,0)$) [rectangle,draw,minimum height=1.5cm] {$P_{Y_1,Y_2|X_1,X_2}$};
	
	\draw[->] (x1) to ($(f1.west)+(0,.5)$);
	\draw[->] (x2) to ($(f1.east)+(0,-.5)$);
	\draw[->] ($(f1.east)+(0,.5)$) to (y2);
	\draw[->] ($(f1.west)+(0,-.5)$) to (y1);
	
	\node (w1) at ($(1.west)-(.8,-.5)$) {$W_1$};
	\node (w2h) at ($(1.west)-(.8,.5)$) {$\hat{W}_2$};
	\node (w2) at ($(2.east)+(.8,-.5)$) {$W_2$};
	\node (w1h) at ($(2.east)+(.8,.5)$) {$\hat{W}_1$};
	\draw[->] (w1) to ($(1.west)-(0,-.5)$);
	\draw[<-] (w2h) to ($(1.west)-(0,.5)$);
	\draw[->] (w2) to ($(2.east)+(0,-.5)$);
	\draw[<-] (w1h) to ($(2.east)+(0,.5)$);
	\end{tikzpicture}
	\caption{A memoryless two-way channel model.}
	\label{fig:two_way}
\end{figure}
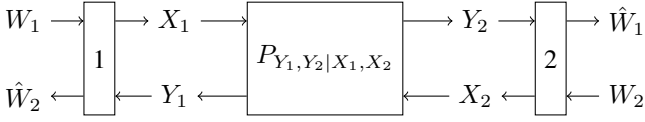

A two-way channel (TWC) has two terminals: each terminal is both a source and a destination. This work considers two different channel settings. One is the canonical discrete-time memoryless channel used in Secs.~\ref{sec:ISD} and \ref{sec:ISD_example} and is completely specified by a transition probability $p(y_1, y_2|x_1, x_2)$ for $x_i \in \mathcal{X}_i$, $y_i \in \mathcal{Y}_i$ for $i \in \{1,2\}$. Herein the alphabet spaces are the real line $\mathbb{R}$, or equivalently finitely discrete after discretization \cite[Chap.~3.4.1]{ElGamalK2011}. The TWC may be subject to cost constraints on inputs. The other is a continuous-time channel whose details are deferred to Sec.~\ref{sec:Poisson}.

The canonical discrete-time TWC, depicted in Fig.~\ref{fig:two_way}, is described as follows: Terminal $i \in \{1, 2\}$ wants to send a message $W_i$ to terminal $j \ne i$, and the messages are independent and uniformly distributed over $\mathcal{W}_i=\{1, \ldots, 2^{nR_i}\}$. Communication takes place over $n$ time units, where in time $t$, terminal $i$ uses an encoding function $\mathcal{E}_{i,t}:\mathcal{W}_i \times \mathcal{Y}_i^{t-1} \mapsto \mathcal{X}_i$ to obtain $X_i(t)=\mathcal{E}_{i,t}(W_i,Y_i(1),\ldots,Y_i(t-1))$ which is sent to terminal $j$. This type of encoding is known as adaptive encoding, contrary to non-adaptive encoding where $X_i(t)=\mathcal{E}_{i,t}(W_i)$. In addition, terminal $i$ uses a decoding function $\mathcal{D}_i: \mathcal{W}_i \times \mathcal{Y}_i^n \mapsto \mathcal{W}_j$ to decode $\hat{W}_j=\mathcal{D}_i(W_i,Y_i(1),\ldots,Y_i(n))$.

The collection of the message sets, encoders, and decoders is known as a code for the TWC. Individual error probabilities $P_{e,1}^{(n)}, P_{e,2}^{(n)}$ are defined as follow.
\begin{align*}
P_{e,1}^{(n)} &= \frac{1}{2^{n(R_1+R_2)}} \sum_{w_1=1}^{2^{nR_1}} \sum_{w_2=1}^{2^{nR_2}} \mathbb{P}[\hat{w}_2 \neq w_2 | w_1, w_2 \textrm{ sent}], \\
P_{e,2}^{(n)} &= \frac{1}{2^{n(R_1+R_2)}} \sum_{w_1=1}^{2^{nR_1}} \sum_{w_2=1}^{2^{nR_2}} \mathbb{P}[\hat{w}_1 \neq w_1 | w_1, w_2 \textrm{ sent}],
\end{align*}
and their maximum is $P_e^{(n)} = \max \left(P_{e,1}^{(n)}, P_{e,2}^{(n)}\right)$. The rate pair $(R_1, R_2)$ is said to be \textit{achievable} if there exists a sequence of codes of rate $(R_1, R_2)$ such that $P_{e}^{(n)} \to 0$ as $n \to \infty$. The closure of the convex hull of the set of all achievable rate pairs $(R_1, R_2)$ is the capacity region $\mathcal{C}$ of the channel, and is the subject of investigation of this paper.

\subsection{Shannon's Inner and Outer Bounds} \label{subsec:Shannon_bounds}
Shannon \cite{Shannon1961} established an inner bound and an outer bound of the capacity region $\mathcal{C}$ for the canonical memoryless TWC. These bounds are described next.

Let $X_1$ and $X_2$ be any input random variables (that meet cost constraints) with corresponding output random variables $Y_1$ and $Y_2$ induced by the channel transition probability. Then note that the joint probability distribution of $(X_1, X_2, Y_1, Y_2)$ is given by:
\begin{align*}
&p_{X_1,X_2,Y_1,Y_2}(x_1,x_2,y_1,y_2) \\
&\quad = p_{X_1,X_2}(x_1,x_2) p_{Y_1,Y_2|X_1,X_2}(y_1,y_2|x_1,x_2).
\end{align*}

Define $\mathcal{P}$ as the collection of all joint distributions $p_{X_1,X_2}$ that satisfy the given constraints. Then, the capacity of the TWC is outer-bounded by
\begin{align*}
\mathcal{C} \subseteq \Rout \triangleq \overline{ \bigcup_{p_{X_1,X_2} \in \mathcal{P}} \mathcal{R}(p_{X_1,X_2}) }
\end{align*} 
where the overline is the closure of the region, and $\mathcal{R}(p_{X_1,X_2})$ is the set of $(R_1,R_2) \in \mathbb{R}_+^2$ satisfying
\begin{align*}
R_1 &\leq I(X_1;Y_2|X_2), \\
R_2 &\leq I(X_2;Y_1|X_1),
\end{align*}
with $(X_1,X_2)$ distributed according to $p_{X_1,X_2} \in \mathcal{P}$. 

On the other hand, define $\mathcal{P}_{\sf prod}$ as the collection of all distributions that factor as $p_{X_1,X_2}=p_{X_1}p_{X_2}$, the product of the marginals of $X_1$ and $X_2$, and satisfy the given constraints. Then, the capacity of the TWC is inner-bounded by
\begin{align*}
\mathcal{C} \supseteq \Rin \triangleq \overline{\sf CH} \left(\bigcup_{p_{X_1,X_2} \in \mathcal{P}_{\sf prod}} \mathcal{R}(p_{X_1,X_2}) \right),
\end{align*}
where $\overline{\sf CH}$ is the convex closure of the convex-hull.

Note that the inner bound, defined with a product input distribution, is proven using an achievability scheme where the two terminals transmit symbols without adaptation to received symbols. Therefore, if the two bounds coincide, then the capacity-achieving distributions of $(X_1,X_2)$ (forming the boundary of $\mathcal{C}$) are product distributions and adaptation is not useful. This is important since nonadaptive coding keeps the system simple. In what follows, we prove that Shannon's bounds (asymptotically) coincide for two classes of channels.

\section{Injective Semi-Deterministic TWC} \label{sec:ISD}
This section discusses a class of discrete-time memoryless channels for which Shannon's inner bound $\Rin$ is tight, i.e., adaptation is useless. We first study sufficient conditions for Shannon's inner bound $\Rin$ to match the outer bound $\Rout$, and then define injective semi-deterministic (ISD) TWCs.

\subsection{Conditions for the Optimality of $\Rin$} \label{subsec:Conditions}
Sufficient conditions under which the inner and outer bounds coincide have been given by Shannon \cite{Shannon1961}. While Shannon's conditions are given in terms of the conditional probability $p_{Y_1,Y_2|X_1,X_2}$, we provide conditions in terms of the conditional entropies, which are simpler and suffice for the purposes of this paper.

\begin{thm} \label{thm:cond}
	The bounds $\Rout$ and $\Rin$ coincide if for all $p_{X_1,X_2}=p_{X_2}p_{X_1|X_2}$, the following holds for $i \in \{1,2\}$:
	\begin{enumerate}[leftmargin=3\parindent]
		\item[(C1)] $H(Y_i|X_1,X_2)$ is invariant with respect to $p_{X_1|X_2}$, and
		\item[(C2)] $H(Y_i|X_i)\leq H(\bar{Y}_i|\bar{X}_i)$, where $\bar{Y}_i$ is the channel output corresponding to independent inputs $\bar{X}_1$ and $\bar{X}_2$ distributed according to $p_{\bar{X}_1}$ and $p_{\bar{X}_2}=p_{X_2}$, respectively, for some $p_{\bar{X}_1}$.
	\end{enumerate}
\end{thm}
\begin{IEEEproof}
	To show that the bounds coincide under conditions (C1) and (C2), consider a distribution $p_{X_1,X_2} = p_{X_2} p_{X_1|X_2}$, with mutual information $I(X_1;Y_2|X_2)=H(Y_2|X_2)-H(Y_2|X_1,X_2)$. Under condition (C1), we have
	\begin{align*}
	H(Y_2|X_1,X_2) &= \mathbb{E}_{X_2}\mathbb{E}_{X_1|X_2}\left[H(Y_2|X_1=x_1,X_2=x_2)\right] \\
	&=\mathbb{E}_{\bar{X}_2}\mathbb{E}_{\bar{X}_1} \left[H(\bar{Y}_2|\bar{X}_1=\bar{x}_1,\bar{X}_2=\bar{x}_2)\right] \\
	&=H(\bar{Y}_2|\bar{X}_1,\bar{X}_2),
	\end{align*}
	since $\bar{X}_2$ has the same distribution as $X_2$, and since $H(Y_2|X_1,X_2)$ is invariant with respect to $p_{X_1|X_2}$. Combining this with condition (C2) leads to
	\begin{align*}
	I(X_1;Y_2|X_2)&=H(Y_2|X_2)-H(\bar{Y}_2|\bar{X}_1,\bar{X}_2) \\
	&\leq H(\bar{Y}_2|\bar{X}_2)-H(\bar{Y}_2|\bar{X}_1,\bar{X}_2) \\
	&=I(\bar{X}_1;\bar{Y}_2|\bar{X}_2).
	\end{align*}
	Similarly, $I(X_2;Y_1|X_1)\leq I(\bar{X}_2;\bar{Y}_1|\bar{X}_1)$. Thus, the rates $R_1$ and $R_2$ corresponding to any  distribution $p_{X_1,X_2}$ are jointly maximized by the rates corresponding to a product distribution $p_{\bar{X}_1}p_{\bar{X}_2}$. Hence, $\Rout \subseteq \Rin$, and hence, the two coincide which concludes the proof. 
\end{IEEEproof}

Interestingly, it turns out by \cite[Sec.~II.F]{WengSAL2019} that the pair of conditions (C1) and (C2) is more general than Shannon's symmetry conditions \cite{Shannon1961}.

\subsection{Injective Semi-Deterministic Channels with Input-Independent Noise} \label{subsec:ISD_TWC}

In this subsection, we define a class of channels that satisfy (C1) and (C2), therefore, adaptation is not necessary to achieve the capacity. As we will see in the next section, this class includes many practical channels such as additive exponential family noise, additive Cauchy noise, and even full-duplex channel with input-dependent noise.

\begin{figure}[t]
	\centering
	\begin{tikzpicture}
	\node (1) at (0,0) [rectangle,draw,minimum height=1.5cm] {1};
	\node (x1) at ($(1)+(1,.5)$) {$X_1$};
	\node (y1) at ($(1)+(1,-.5)$) {$Y_1$};
	\draw[->] ($(1.east)+(0,.5)$) to (x1);
	\draw[<-] ($(1.east)+(0,-.5)$) to (y1);
	
	\node (2) at (6,0) [rectangle,draw,minimum height=1.5cm] {2};
	\node (x2) at ($(2)+(-1,-.5)$) {$X_2$};
	\node (y2) at ($(2)+(-1,.5)$) {$Y_2$};
	\draw[->] ($(2.west)+(0,-.5)$) to (x2);
	\draw[<-] ($(2.west)+(0,.5)$) to (y2);
	
	\node (f1) at ($(y1)+(.9,0)$) [rectangle,draw,minimum height=.6cm] {$f_1$};
	\node (g1) at ($(f1)+(1.1,0)$) [rectangle,draw,minimum height=.6cm] {$g_1$};
	\node (f2) at ($(y2)+(-.9,0)$) [rectangle,draw,minimum height=.6cm] {$f_2$};
	\node (g2) at ($(f2)+(-1.1,0)$) [rectangle,draw,minimum height=.6cm] {$g_2$};
	
	\draw[->] (x1) to (g2.west);
	\draw[->] (g2.east) to node[above] {$T_2$} (f2.west);
	\draw[->] (x2) to (g1.east);
	\draw[->] (g1.west) to node[above] {$T_1$} (f1.east);
	\draw[->] (f2) to (y2);
	\draw[->] (f1) to (y1);
	\draw[->] ($(x1)+(.9,0)$) to (f1.north);
	\draw[->] ($(x2)+(-.9,0)$) to (f2.south);
	
	\node (z2) at ($(g2)+(0,.8)$) {$Z_2$};
	\node (z1) at ($(g1)+(0,-.8)$) {$Z_1$};
	\draw[->] (z2) to (g2.north);
	\draw[->] (z1) to (g1.south);
	\end{tikzpicture}
	\caption{A semi-deterministic two-way channel. With conditions that $f_i, g_i$ are injective, it is an ISD TWC.}
	\label{fig:SDTWC}
\end{figure}
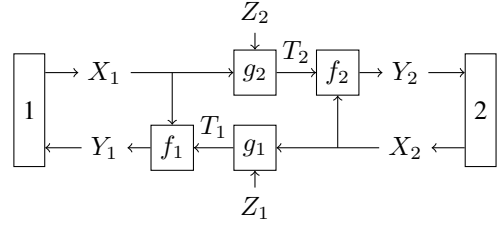

Let us define two functions for $i,j\in\{1,2\}$, $i\neq j$,
\begin{align*}
g_i:\mathcal{X}_j \times \mathcal{Z}_i \mapsto \mathcal T_i, \text{ and } f_i:\mathcal{X}_i \times \mathcal{T}_i \mapsto \mathcal{Y}_i,
\end{align*}
for some sets $\mathcal{Z}_i$ and $\mathcal{T}_i$. Further, assume that $f_i(X_i,T_i)$ is injective in $T_i$, i.e., for every $x_i\in\mathcal{X}_i$, $f_i(x_i,t_i)$ is one-to-one in $t_i\in\mathcal{T}_i$, and that $g_i(X_j,Z_i)$ is injective in $Z_i$, i.e., for every $x_j\in\mathcal{X}_j$, $g_i(x_j,z_i)$ is one-to-one in $z_i\in\mathcal{Z}_i$.\footnote{Related conditions for one-way channels with feedback were given in \cite{AlajajiF1994}.} We define an ISD TWC with input-independent noise, illustrated in Fig.~\ref{fig:SDTWC}, as follows.

\begin{defi}[ISD TWC] \label{def:ISD_TWC}
	The injective semi-deterministic TWC with input-independent noise is one with
	\begin{align*}
		Y_i=f_i(X_i,T_i), \quad \text{and} \quad T_i=g_i(X_j,Z_i),
	\end{align*}
	$i,j\in\{1,2\}$, $i\neq j$, where $Z_1\in\mathcal{Z}_1$ and $Z_2\in\mathcal{Z}_2$ are (possibly dependent on each other) random variables independent of $X_1$ and $X_2$.
\end{defi}

For this class, we can prove the following capacity result.
\begin{thm} \label{thm:ISDTWC}
	For the class of ISD TWCs with input-independent noise, conditions (C1) and (C2) are satisfied and the capacity region is $\mathcal{C}=\Rin = \Rout$. 
\end{thm}
\begin{IEEEproof}
	By Thm.~\ref{thm:cond}, it is sufficient to show that conditions (C1) and (C2) are satisfied. We have
	\begin{align*}
	H(Y_1|X_1,X_2)&=H(f_1(X_1,g_1(X_2,Z_1))|X_1,X_2) \\
	&=H(g_1(X_2,Z_1)|X_1,X_2) \\
	&=H(Z_1|X_1,X_2) \\
	&=H(Z_1),
	\end{align*}
	which follows from the injectivity of $f_1$ and $g_1$, and the independence of $Z_1$ and $(X_1,X_2)$. This is also independent of $p_{X_1|X_2}$. Similarly, $H(Y_2|X_1,X_2)=H(Z_2)$, and hence, (C1) is satisfied. On the other hand, 
	\begin{align*}
	H(Y_1|X_1)&=H(f_1(X_1,g_1(X_2,Z_1))|X_1) \\
	&=H(g_1(X_2,Z_1)|X_1) \\
	&\leq H(g_1(X_2,Z_1)),
	\end{align*}
	again by the injectivity of $f_1$. This upper bound is equal to the entropy of the output $Y_1$ when the inputs are independent.
	
	Similarly, $H(Y_2|X_2)$ is upper-bounded by $H(g_2(X_1,Z_2))$, the entropy of $Y_2$ when $X_1$ and $X_2$ are independent. Thus, (C2) is also satisfied, which implies that the bounds $\Rin$ and $\Rout$ coincide by Thm.~\ref{thm:cond} and the statement follows.
\end{IEEEproof}

As a corollary of Thm.~\ref{thm:ISDTWC}, we have the following.
\begin{cor} \label{cor:rect}
	For the class of ISD TWCs with input-independent noise, the capacity region is rectangular given by the set of $(R_1,R_2)\in\mathbb{R}_+^2$ so that, for $i,j\in\{1,2\}$, $i\neq j$:
	\begin{align*}
	R_i\leq \max_{p_{X_i}} H(g_j(X_i,Z_j)) - H(Z_j).
	\end{align*}
\end{cor}
\begin{IEEEproof}
	The proof follows directly from the fact that $H(Y_i|X_1,X_2)$ is independent of $p_{X_1,X_2}$, and that $H(Y_i|X_i)\leq H(g_i(X_j,Z_i))$ which is maximized by the same $p_{X_j}$ independent of $p_{X_i}$, for $i,j\in\{1,2\}$, $i\neq j$.
\end{IEEEproof}

\section{Examples} \label{sec:ISD_example}
Thm.~\ref{thm:ISDTWC} for the ISD TWCs enables us to characterize the capacity of many classical TWCs in closed form. They include multiplicative channels, additive channels with various types of input-independent noises, and also input-dependent ones.

\subsection{Multiplicative TWC}
The class in Def.~\ref{def:ISD_TWC} subsumes other channels such as multiplicative channels with input-independent noise, where $Y_i=X_i\cdot X_j\cdot Z_i$ such that $0 \notin \mathcal{X}_i\cup\mathcal{Z}_i$. In this case, $f_i(X_i,T_i) =X_i\cdot T_i$ and $g_i(X_j,Z_i)=X_j\cdot Z_i$, which are injective given $X_i$ and $X_j$, respectively. This leads to the following corollary.
\begin{cor} \label{cor:multiplicative_TWC}
	The capacity region of a multiplicative TWC with input-independent noise, where $Y_i=X_i\cdot X_j\cdot Z_i$, $i\in\{1,2\}$, $i\neq j$, so that $0\not\in\mathcal{X}_i\cup\mathcal{Z}_i$, is given by the rectangle defined by 
	\begin{align}
	0\leq R_i\leq \max_{p_{X_i}}I(X_i;X_i\cdot Z_j).
	\end{align}
\end{cor}
The binary multiplier channel \cite{Shannon1961} for which the bounds do not coincide does not belong to this class because $0\in\mathcal{X}_i$ (contrary to Cor.~\ref{cor:multiplicative_TWC}), and thus, the channel is not injective. Capacity outer bounds for this channel, which are tighter than Shannon's, had been given in \cite{HekstraW1989}.

\subsection{Additive TWC with Input-Independent Noise}
The additive two-way channel is defined by the channel mapping:
\begin{align*}
Y_1 &= X_1 + X_2 + Z_1, \\
Y_2 &= X_1 + X_2 + Z_2,
\end{align*}
where $X_1, X_2$ are channel inputs satisfying given constraints, and $Z_1, Z_2$ are independent of $X_1, X_2$ but, possibly dependent on each other. Obviously it is ISD by taking $f_i(X_i, T_i) = X_i+T_i$ and $g_i(X_j, Z_i) = X_j+Z_i$ so the capacity region is rectangular by Cor.~\ref{cor:rect}. Furthermore, for some classes, we can characterize the capacity region expressions in closed form.

\subsubsection{Exponential Noise}
Suppose $Z_1$ and $Z_2$ are exponential random variables with means $m_1$ and $m_2$ respectively. Further there are expected amplitude constraints $A_1$ and $A_2$ on the channel input sequences $\{x_1(t)\}_{t=1}^n$ and $\{x_2(t)\}_{t=1}^n$, respectively: i.e., for all message $w_1 \in \mathcal{W}_1$, $w_2 \in \mathcal{W}_2$,
\begin{equation} \label{eq:amp_constraint}
\mathbb{E} \left[ \sum_{t=1}^n x_1(t) \right] \le nA_1 ~ \textrm{ and } ~ \mathbb{E} \left[ \sum_{t=1}^n x_2(t) \right] \le nA_2.
\end{equation}

The capacity expression in closed form is as follows.
\begin{thm} \label{thm:AWEN_cap}
	For the additive exponential noise channel, the capacity region is given by
	\begin{align*}
	\mathcal{C}(A_1,A_2) = \begin{cases}
	R_1 \le \log\left(1 + \frac{A_2}{m_1}\right), \\
	R_2 \le \log\left(1 + \frac{A_1}{m_2}\right).
	\end{cases}
	\end{align*}
\end{thm}
\begin{IEEEproof}
	Since this TWC is ISD, i.e., $\mathcal{C} = \Rin$, it suffices to characterize the bounds in $\Rin$. Note that independent one-way capacity-achieving inputs $X_i$ that meet the amplitude constraint $\mathbb{E}[X_i] \le A_i$ are given by the saddle-point result \cite{Verdu1996},\cite[Thm.~3]{AnantharamV1996},\cite[Thm.~1]{ColemanR2010},
	\begin{align*}
	p_{X_i}(x) = \frac{m_i}{A_i+m_i} \delta(x) +\frac{A_i}{A_i+m_i} e^{-\frac{x}{m_i + A_i}}, ~~ i=1,2,
	\end{align*}
	where $\delta(x)$ is the Dirac-delta function, i.e., a point mass at $0$.
	
	Then since the effect of $X_2$ on $Y_2$ can be subtracted at terminal $2$ and by the one-way saddle-point computation \cite{Verdu1996},
	\begin{align*}
	I(X_1; Y_2 | X_2) = \log\left(1 + \tfrac{A_1}{m_2}\right).
	\end{align*}
	Likewise, 
	\begin{align*}
	I(X_2; Y_1 | X_1) = \log\left(1 + \tfrac{A_2}{m_1}\right).
	\end{align*}
	This completes the characterization.
\end{IEEEproof}

Results on additive general exponential family noise with input cost function $\rho: \mathbb{R} \mapsto \mathbb{R}^+$, which subsumes Gaussian noise with $\rho(x) = x^2$ \cite{Han1984}, can be found in \cite{Varshney2013c} and are omitted for brevity.

\subsubsection{Cauchy Noise}
Suppose $Z_1, Z_2$ are Cauchy, i.e., 
\begin{align*}
p_{Z_i}(z_i)=\left[ \pi\gamma_i\left( 1 + \left( \frac{z_i}{\gamma_i} \right)^2 \right) \right]^{-1},
\end{align*}
where $\gamma_i$ is the dispersion parameter and the location parameter is assumed to be zero without loss of generality. This is henceforth denoted $P_{\sf Cauchy}(0,\gamma_i)$. The Cauchy distribution models impulsive noise or interference caused by Poisson distributed interferers, where the second moment of noise can be infinite and not suitable as a power measure \cite{FahsA2014}. In this case, it is common to assign a logarithmic constraint
\begin{align} \label{eq:Cauchy_constraint}
C(A_i,\gamma_i) \triangleq \mathbb{E}_{X} \left[ \log \left( \left( \frac{A_i+\gamma_i}{A_i}\right)^2 + \left(\frac{X}{A_i}\right)^2 \right) \right] \leq \log(4),
\end{align}
for some $A_i \geq \gamma_i$, viewed as a power constraint. Then, the capacity expression in closed form is as follows.

\begin{thm}
	For the Cauchy TWC, the capacity region is given by
	\begin{align*}
	\mathcal{C}(A_1,A_2) = \begin{cases}
	R_1 \le \log\left(\frac{A_1}{m_2}\right), \\
	R_2 \le \log\left(\frac{A_2}{m_1}\right),
	\end{cases}
	\end{align*}
	and is achieved by $X_i$ distributed according to $P_{\sf Cauchy}(0,A_i-\gamma_j)$.
\end{thm}
\begin{IEEEproof}
	Since this channel belongs to the ISD class, its capacity is given by $\Rin$, described by $0\leq R_i\leq \max_{P_{X_i}}h(Y_j)-h(Z_j)$, $i\neq j$ by Cor.~\ref{cor:rect}. The maximization here is subject to \eqref{eq:Cauchy_constraint}. Suppose input $X_j$ satisfies
	\begin{align} \label{eq:Cauchy_constraint2}
	\mathbb{E}_{X_j}\left[\log\left(\left(\frac{k_j+\gamma_i}{k_j}\right)^2+\left(\frac{X_j}{k_j}\right)^2\right)\right]= \log(4),
	\end{align}
	with some $k_j\in[\gamma_i,A_j]$, which is an equivalent expression to \eqref{eq:Cauchy_constraint} \cite{FahsA2014}. Then, by \cite{FahsA2014},
	\begin{align}
	\mathbb{E}_{Y_i}\left[\log\left(1+\left(\frac{Y_i}{k_j}\right)^2\right)\right]= \log(4). \label{eq:Cauchy_constraint3}
	\end{align}
	Under constraints \eqref{eq:Cauchy_constraint2} and \eqref{eq:Cauchy_constraint3}, we solve $\max_{P_{X_i}}h(Y_j)-h(Z_j)$: the entropy of $Y_i$ is maximized when $Y_i$ is distributed according to $P_{\sf Cauchy}(0,k_j)$, which can be attained if $X_j$ is distributed according to $P_{\sf Cauchy}(0,k_j-\gamma_i)$ and satisfies \eqref{eq:Cauchy_constraint2}. The entropy of a Cauchy random variable $P_{\sf Cauchy}(0,\mu)$ is $\log(4\pi\mu)$. Thus, the rate $R_j$ satisfies
	\begin{align*}
	R_j &\leq \max_{P_{X_j}}h(Y_i)-h(Z_i)\\
	&= h(Y_i)|_{X_j\sim P_{\sf Cauchy}(0,k_j-\gamma_i)}-h(Z_i)\\
	&= \log\left(\frac{k_j}{\gamma_i}\right).
	\end{align*}
	Since this is increasing in $k_j$, the fact that the maximum is at $k_j=A_j$ leads to the desired result.
\end{IEEEproof}

This channel provides an interesting example of an ISD TWC not of the exponential family, where adaptation is not necessary. A key element is the independence between the noise and the inputs. Dependence, however, does not imply the necessity of adaptation as explained next.

\subsection{Relaxation: Input-Dependent Noise}
While channels with input-dependent noise do not belong to the class defined in Def.~\ref{def:ISD_TWC}, they might still satisfy (C1) and (C2). An example is the channel in \cite[Tab.~II]{Shannon1961} which is ISD but with input-dependent noise. Namely, in this example, the inputs and outputs are binary, with
\begin{align*}
Y_1 = X_2, \quad \text{ and } \quad Y_2 = \begin{cases}X_1 &\text{if } X_2=0,\\ N_2&\text{if } X_2=1,\end{cases}
\end{align*} 
where $N_2\in\{0,1\}$ is Bern$(1/2)$ (Bernoulli distributed). This can be modeled as an ISD TWC where 
\begin{align*}
g_1(X_2,Z_1)&=X_2, &g_2(X_1,Z_2)&=X_1+Z_2,\\
f_1(T_1,X_1)&=T_1, &f_2(X_2,T_2)&=T_2,
\end{align*}
$\mathcal{Z}_1=\{0\}$, $\mathcal{Z}_2=\{0,1\}$, $Z_2=X_2(N_2+X_1)$, and where addition is modulo-$2$. Here, noise is input-dependent, yet, (C1) and (C2) are satisfied, and adaptation is not necessary.

Another example of this sort with practical relevance is an additive TWC with input-dependent Gaussian noise, with
\begin{align*}
Y_i = a_iX_i + X_j + \sqrt{X_j} \tilde{Z}_i + \hat{Z}_i,
\end{align*}
$i,j \in \{1,2\}$, $i\neq j$, where $X_i \in \mathbb{R}_+$, $\hat{Z}_i, \tilde{Z}_i \in \mathbb{R}$, $\hat{Z}_i$ and $\tilde{Z}_i$ are Gaussian noises with zero mean and variances $\hat{\sigma}_i^2$ and $\tilde{\sigma}_i^2$, respectively, independent of $X_1$ and $X_2$, and $X_i$ satisfies cost constraint $\mathbb{E}[c(X_i)]\leq P_i$ for some cost function $c(\cdot)$. This channel is not covered by Han's result \cite{Han1984} where input-independent noise was assumed. Such an input-dependent noise model is introduced in \cite{Moser2012} to model optical wireless communications.\footnote{This model assumes an intensity detector. A model with a photon detector corresponds to the Poisson TWC in Sec.~\ref{sec:Poisson}.} This channel can be modeled as an ISD TWC with $Z_i=\sqrt{X_j}\tilde{Z}_i+\hat{Z}_i$, and $Y_i=f_i(X_i,T_i)=a_iX_i+T_i$ and $T_i=g_i(X_j,Z_i)=X_j+Z_i$ which are injective in $T_i$ and $Z_i$ respectively. The main difference with Def.~\ref{def:ISD_TWC} is that noise $Z_i$ is input-dependent. Nevertheless, the capacity region of this channel can be determined. In particular,
\begin{align*}
&I(X_1;Y_2|X_2) \\
&=h(Y_2|X_2) - h(\sqrt{X_1} \tilde{Z}_2 + \hat{Z}_2|X_1)\\
&= h(X_1 + \sqrt{X_1}\tilde{Z}_2 + \hat{Z}_2) - h(\sqrt{X_1}\tilde{Z}_2 + \hat{Z}_2|X_1) \\
&\leq \max_{P_{X_1}} \left[ h(X_1 + \sqrt{X_1}\tilde{Z}_2 + \hat{Z}_2)\right. \\
&\hspace{1.8cm} \left. - \mathbb{E}_{X_1} \left[ \frac{1}{2} \log(2\pi e(X_1 \tilde{\sigma}_2^2 + \hat{\sigma}_2^2)) \right] \right] \triangleq \bar{C}_1,
\end{align*}
and similarly $I(X_2;Y_1|X_1)\leq\bar{C}_2$ with $\bar{C}_2$ defined similarly with interchanged indices ($1\leftrightarrow2$). Noting that $\bar{C}_i$ does not depends on $X_j$, these upper bounds are achievable using independent maximization in $X_1$ and $X_2$. Thus, the capacity of this channel is given by the rectangular region defined by
\begin{align*}
0\leq R_1&\leq \bar{C}_1,\text{ and }0\leq R_2\leq \bar{C}_2,
\end{align*}
which is achievable without adaptation.

\section{Poisson TWC} \label{sec:Poisson}
Now we turn our attention to a continuous-time TWC that was not studied in our previous conference papers \cite{Varshney2013c, ChaabanVA2017}, but that is also of practical importance for full-duplex communication, cf.\ \cite{GhassemlooyPR2018}. Note that the Poisson TWC is a continuous-time channel unlike discrete-time channels in Secs.~\ref{sec:ISD} and \ref{sec:ISD_example}. However, Wyner \cite{Wyner1988a, Wyner1988b} simplified the one-way Poisson channel into a sequence of i.i.d.~discrete-time channels with the capacity unchanged.

\subsection{Channel Model and Definition}
The Poisson channel has been used to model pulse-amplitude modulated optical communication. The one-way Poisson channel is described as follows: The transmitter modulates the input current waveform $\lambda(t), t \in [0, T]$ of a light source in continuous-time. The waveform has peak power constraint $A$, i.e., $\lambda(t) \in [0, A]$, and average power constraint with given $\sigma \in [0,1]$,
\begin{align*}
\frac{1}{T} \int_0^{T} \lambda(t) dt \le \sigma A, \quad \sigma \in [0,1].
\end{align*}
Then, the receiver detects photons by a continuous-time Poisson process $\Phi_Y(t) = \Phi_X(t;\lambda(t))+\Phi_Z(t;\lambda_0)$, where $\Phi(t;\lambda(t))$ denotes a Poisson point process (or equivalently, a counting process) with instantaneous rate $\lambda(t)$, and $\lambda_0$ is the dark current intensity. The receiver has a direct detector of photons, i.e., the receiver observes photons over time.

A codebook of size $M$ for the Poisson channel is a set of $M$ waveforms over $T$, $\{\lambda^{(1)}, \ldots, \lambda^{(M)}\}$, where each waveform satisfies the peak power and average power constraints mentioned above. The average error probability is 
\begin{align*}
P_e = \frac{1}{M} \sum_{m=1}^M \Pr\left[ \mathcal{D}(\Phi_Y(t), 0 \le t \le T) \ne m | \lambda^{(m)} \textrm{ was sent}\right],
\end{align*}
where $\mathcal{D}(\cdot)$ is a decoding function. Then a rate $R \ge 0$ is said to be \textit{achievable} if for any $\epsilon > 0$, there exists a codebook of size $\frac{1}{T}\log M \ge R$ ($T$ can be arbitrarily large) and a decoder such that $P_e \le \epsilon$. The \textit{capacity} is the supremum of achievable rates. Note that the unit of the capacity is nats per unit time.

Similarly, a Poisson TWC is defined as
\begin{align*}
\Phi_{Y_1}(t) &= \Phi_{X_1}(t;\lambda_1(t)) + \Phi_{X_2}(t;\lambda_2(t)) + \Phi_{Z_1}(t;\lambda_0), \\
\Phi_{Y_2}(t) &= \Phi_{X_1}(t;\lambda_1(t)) + \Phi_{X_2}(t;\lambda_2(t)) + \Phi_{Z_2}(t;\lambda_0),
\end{align*}
where input intensities satisfy peak power constraint $A$ and average power constraint $\sigma_1, \sigma_2$, i.e.,
\begin{align*}
\frac{1}{T} \int_0^{T} \lambda_i(t) dt \le \sigma_i A,
\end{align*}
and $\Phi_{Z_1}, \Phi_{Z_2}$ are dark current noises, independent of other random variables. We assume the same peak power constraints and dark current intensities of both inputs for mathematical brevity. Generalization does not significantly change results of this paper.

The capacity region of the Poisson TWC is defined in the same way: a rate pair $(R_1, R_2)$ is said to be achievable if for any $\epsilon > 0$, there exists a pair of codebooks of size $\frac{1}{T} \log M_i \ge R_i$ such that $P_{e,i} \le \epsilon$. The union of all such rate pairs forms the capacity region, $\mathcal{C}(\sigma_1, \sigma_2)$, provided that $A, \lambda_0$ are given.

\subsection{Equivalent Discrete-Time Binary TWC}
As stated, Wyner \cite{Wyner1988a, Wyner1988b} converted the Poisson channel, a continuous-time channel, into a sequence of i.i.d.~discrete-time channels. The next lemma for the Poisson TWC, immediate from the one-way Poisson channel result \cite{Wyner1988a, Wyner1988b}, also gives the same conclusion.

\begin{lem} \label{lem:Poisson_to_DMC}
	For any code of rate $(R_1, R_2) \in \mathcal{C}$ of the Poisson TWC, there exists a new code that achieves the same rate pair and satisfies the following properties, provided that $n$ is large enough and $\Delta = \frac{T}{n} > 0$ is sufficiently small.
	\begin{itemize}
		\item The transmitted signal $\lambda_i(t)$ is constant on the interval $[k\Delta, (k+1)\Delta)$ for every $k \in \mathbb{Z}_+$. Furthermore, $\lambda_i(t)$ only takes the values $0$ or $A$.
		
		\item The receiver declares $1$ when it observes one photons on each interval $[k\Delta, (k+1)\Delta)$, and declares $0$ otherwise.
	\end{itemize}
\end{lem}
\begin{IEEEproof}[Proof Sketch]
Wyner's three-step approximation for the one-way channel \cite{Wyner1988b} applies without significant changes.

1) Let $\nu_i(t)$ be the counting process corresponding to $\Phi_{Y_i}(t)$. Then there exists a discrete-time decoder $\hat{\mathcal{D}}_i:\{ \nu(n\delta) \}_{n=1}^{N'} \mapsto \{1,\ldots, M_j\}$ that shows almost identical probability of error without significant increase.

2) Note that $\hat{\mathcal{D}_i}$ only depends on the number of photons over intervals $[(j-1)\delta, j\delta), j=1,\ldots N'$. Then, for a given number of photons, we can arbitrarily control the waveform on each interval. Take the equivalent waveform $\tilde{\lambda}_i(t)$ such that
\begin{align*}
\tilde{\lambda}_i(t) = \begin{cases}
A & ~ (j-1)\delta \le t < t_{mj} \\
0 & ~ t_{mj} \le t < j\delta
\end{cases} ~~~ \textrm{for some } t_{mj}.
\end{align*}

3) Note that $t_{mj}$ is not necessarily at the interval boundary. Split each interval of length $\delta$ into $L$ subintervals, and approximate $\tilde{\lambda}_i(t)$ by the waveform $\hat{\lambda}_i(t)$ that has a transition only at the subinterval boundary. Then the error probability loss is negligible if $L$ is large enough.

Therefore the first claim has been proved and the second claim follows since when $\Delta$ is small enough the probability that there are more than one photons is negligible by the Poisson law.
\end{IEEEproof}

Due to Lem.~\ref{lem:Poisson_to_DMC}, we can induce the equivalent discrete-time binary alphabet TWC using Poisson process law on a small interval. The channel input $X_i$ is $0$ if $\lambda_i(t) = 0$ on the interval $[k\Delta, (k+1)\Delta)$, and $1$ if $\lambda_i(t)=A$.

Note that the number of photons on a small interval of length $\Delta$ emitted by $\Phi(t;\lambda)$ is Poisson distributed with parameter $\lambda \Delta$, so dominant events are either no photon or one photon when $\Delta$ is small enough. Hence we can say the output $Y_i$ is the number of photons observed on an interval of length $\Delta$, thus binary.

Recalling the Poisson distribution with parameter $\lambda \Delta$,
\begin{align*}
&\mathbb{P}[\textrm{one photon on the interval}] = \lambda \Delta \exp(-\lambda \Delta).
\end{align*}
Hence, the discrete-time channel induced by Lem.~\ref{lem:Poisson_to_DMC} is
\begin{align*}
W_i(Y_i|X_1, X_2) &= \kbordermatrix{
	& Y_i=0 & Y_i=1 \\
	X_1X_2 = 00 & 1-\alpha & \alpha \\
	X_1X_2 = 01 & 1-\beta & \beta \\
	X_1X_2 = 10 & 1-\beta & \beta \\
	X_1X_2 = 11 & 1-\gamma & \gamma
},
\end{align*}
where
\begin{align*}
\alpha &\triangleq \lambda_0 \Delta \exp(-\lambda_0 \Delta), \\
\beta &\triangleq (A+\lambda_0) \Delta \exp(-(A+\lambda_0) \Delta), \\
\gamma &\triangleq (2A+\lambda_0) \Delta \exp(-(2A+\lambda_0) \Delta).
\end{align*}

Letting $s = \lambda_0/A$, $W_i$ can be approximated at small $\Delta$ using the first terms of the Taylor series expansion\footnote{Thus, $s \to \infty$ implies the high dark current regime.}:
\begin{align} \label{eq:DMC_matrix}
W_i(Y_i|X_1, X_2) &= \begin{bmatrix}
1-As\Delta & As\Delta \\
1-A(1+s)\Delta & A(1+s)\Delta \\
1-A(1+s)\Delta & A(1+s)\Delta \\
1-A(2+s)\Delta & A(2+s)\Delta
\end{bmatrix}.
\end{align}

\subsection{Poisson TWC Capacity}
Before proceeding to the capacity of the Poisson TWC, we review the capacity of the one-way Poisson channel \cite{Davis1980, Wyner1988a, Wyner1988b}.
\begin{thm}[\cite{Davis1980, Wyner1988a, Wyner1988b}] \label{thm:Poisson_owc}
	The capacity of the one-way Poisson channel is:
	\begin{align*}
	\Cowc(\sigma) &= \lim_{\Delta \to 0} \frac{1}{\Delta} \max_{p_X: \mathbb{E}[X] \le \sigma}I(X;Y) \\
	&= A \Big[ \pi^* (1+s) \log(1+s) + (1-\pi^*) s \log (s) \\
	&\qquad - (\pi^* + s)\log (\pi^*+s) \Big] \quad \textrm{[nats per second]},
	\end{align*}
	where
	\begin{align*}
	s &= \lambda_0 / A, \\
	\pi^* &= \min(\sigma, \pi_0(s)),
	\end{align*}
	and
	\begin{align*}
	\pi_0(s) = \frac{(1+s)^{1+s}}{s^s e} - s.
	\end{align*}
	In addition, the capacity is achieved by $p^*$ with $p^*(1) = \pi^*$.
\end{thm}

From Wyner's discretization argument \cite{Wyner1988a, Wyner1988b}, we can derive Shannon's inner bound for the TWC. We can derive an outer bound using one-way capacities, which is simpler than Shannon's outer bound, but sufficient for the purpose of this paper. These inner and outer bounds are given next.
\begin{defi}[Shannon inner bound] \label{def:Shannon_inner_Poisson}
	Let $\Rin$ be the set of all rate pairs $(R_1, R_2)$ that satisfy
	\begin{align*}
	R_1 &\le \lim_{\Delta \to 0} \frac{1}{\Delta} I(X_1;Y_2|X_2), \\
	R_2 &\le \lim_{\Delta \to 0} \frac{1}{\Delta} I(X_2;Y_1|X_1),
	\end{align*}
	over all distributions $p(x_1)p(x_2)$ satisfying input constraints
	\begin{align*}
		\mathbb{E}[X_i] \le \sigma_i.
	\end{align*}
\end{defi}
\begin{defi}[One-way outer bound] \label{def:onw_way_outer_poisson}
	Let $\Routt$ be the set of all rate pairs $(R_1, R_2)$ that satisfy
	\begin{align*}
	R_1 &\le \Cowc(\sigma_1), \\
	R_2 &\le \Cowc(\sigma_2).
	\end{align*}
	In other words, $\Routt$ is the rectangular region, in which the range of each rate is determined by $\Cowc(\sigma_i)$.
\end{defi}

Then, the following proposition is immediate.
\begin{prop}
	$\Rin \subset \mathcal{C}(\sigma_1, \sigma_2) \subset \Routt$.
\end{prop}
\begin{IEEEproof}
	First inclusion $\Rin \subset \mathcal{C}(\sigma_1, \sigma_2)$ is obvious from Wyner's discretization argument \cite{Wyner1988a, Wyner1988b} and Shannon's inner bound. Second inclusion $\mathcal{C}(\sigma_1, \sigma_2) \subset \Routt$ can be obtained by introducing a genie that tells terminal $i$ the exact number of photons emitted by itself. So the channel is effectively a pair of independent one-way channels and $\mathcal{C}(\sigma_1, \sigma_2) \subset \Routt$ holds.
\end{IEEEproof}

Note from the asymptotic rate of Thm.~\ref{thm:Poisson_owc} that when $s \to \infty$, $\Routt$ vanishes as $A\pi_i^*(1-\pi_i^*)/2s$ along each coordinate in large dark current regime, where $\pi_i^* = \min(\sigma_i, 1/2)$. Hence, we can at least say $\mathcal{C}(\sigma_1, \sigma_2)$ shrinks at $O(s^{-1}) = O(\lambda_0^{-1})$ since $\Routt$ is an outer bound. The next theorem shows that the gap between $\Rin$ and $\Routt$ shrinks faster, i.e., the capacity region is asymptotically rectangular.

\begin{figure}[t]
	\centering
	\includegraphics[width=3.5in]{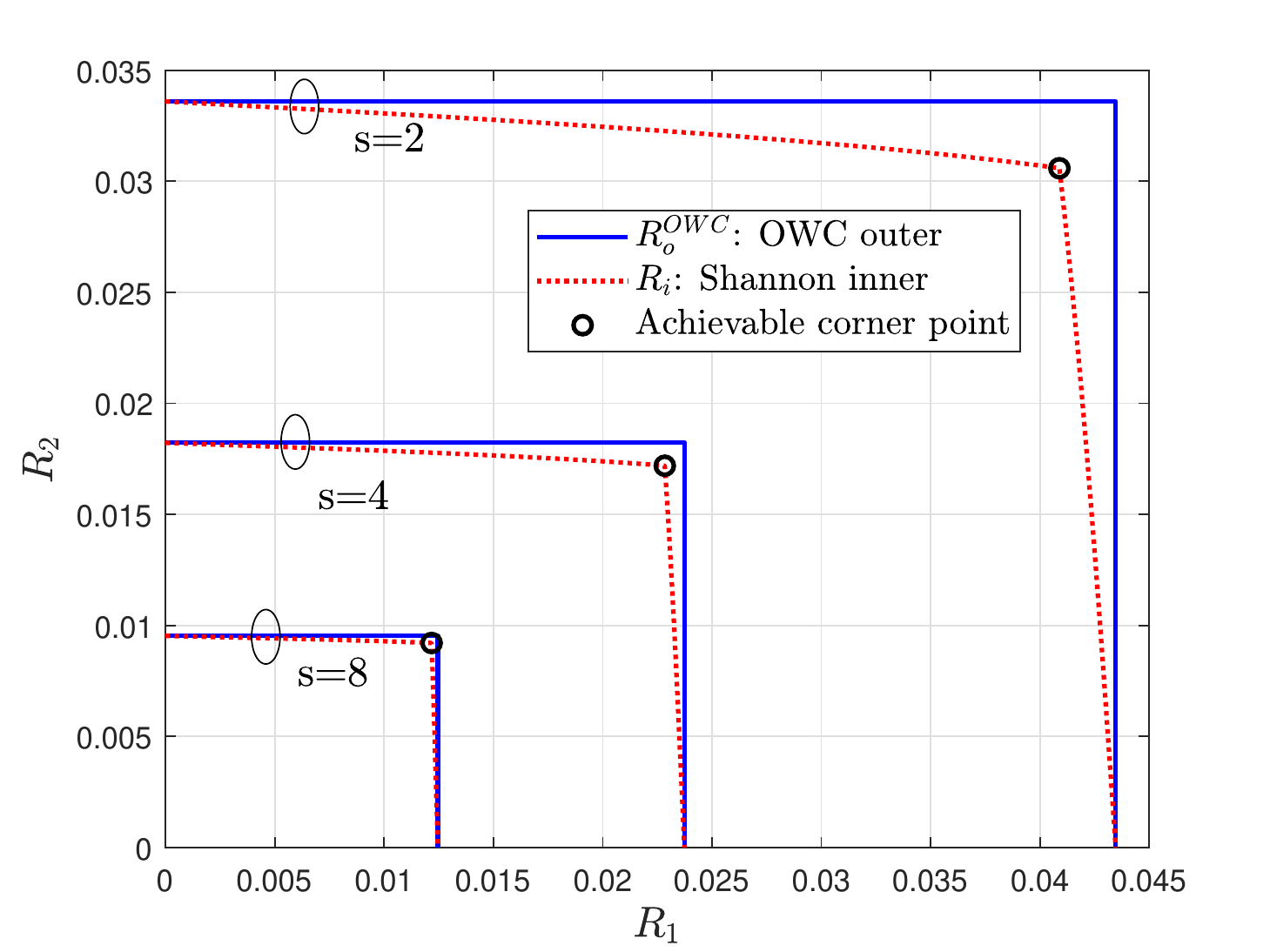}
	\caption{Inner and outer bounds for the Poisson TWC with $A=1, \sigma_1=0.3, \sigma_2=0.2, \Delta=0.0001$. The dark current intensity ranges in $\{2, 4, 8\}$. Circles indicate the corner points achievable by $(
		p_1^*, p_2^*)$.}
	\label{fig:Poisson_plot}
\end{figure}

\begin{thm}\label{thm:gap_vanish}
	$\Rin$ and $\Routt$ asymptotically meet when $s \to \infty$. Furthermore, the gap between boundaries of $\mathcal{C}(\sigma_1, \sigma_2)$ and $\Routt$ vanishes as $O(s^{-2})=O(\lambda_0^{-2})$.
\end{thm}
\begin{IEEEproof}
	Provided in Appendix.
\end{IEEEproof}
Fig.~\ref{fig:Poisson_plot} depicts $\Rin$ in Def.~\ref{def:Shannon_inner_Poisson} and $\Routt$ in Def.~\ref{def:onw_way_outer_poisson} and Thm.~\ref{thm:Poisson_owc} at specified parameters. We can observe that the overall regions shrink at $O(s^{-1})$ as $s$ becomes greater, i.e., as the dark current becomes larger. In addition, the corner points achieved by $(p_1^*, p_2^*)$, where $p_i^*$ is the capacity-achieving distribution of $\Cowc(\sigma_i)$, asymptotically meet the corner point of the one-way outer bound at speed $O(s^{-2})$.

\section{Discussion and Open Problem}
We have studied two classes of full-duplex TWC models that do not require adaptation to achieve capacity regions. The first is the injective semi-deterministic for which Shannon's inner bound meets Shannon's outer bound. The second is the Poisson TWC for which non-adaptive transmission asymptotically achieves the capacity region when the dark current intensity goes to infinity.

A major difficulty in full-duplex implementation is due to the residual self-interference \cite{ChoiJSLK2010}. Since the signal from a local antenna is much stronger than signals from remote antennas, even a small error in echo-channel estimation results in a huge interference. One appealing model for such residual self-interference is an extension of the AWGN TWC,
\begin{align*}
Y_1 &= c_{11} X_1 + \tilde{Z}_1 X_1 + c_{12} X_2 + Z_1, \\
Y_2 &= c_{21} X_1 + \tilde{Z}_2 X_2 + c_{22} X_2 + Z_2,
\end{align*}
where $\tilde{Z}_i$ is Gaussian distributed and models the echo-channel estimation error \cite{LiuSSTG2017}. Interestingly, the capacity region of this model is still open: It does not belong to the ISD TWC or any capacity-known TWC classes, e.g., \cite{WengSAL2019}. Characterizing the capacity region or investigating whether adaptation is useful in this TWC model is an interesting topic for future research.

Moving beyond TWCs, we can also consider multi-way networks. For example, adaptation is useless in some models \cite{ChengD2014}. For the multiple-input and multiple-output setting, it is also known that non-adaptive transmission achieves optimal degrees-of-freedom in some multi-way networks \cite[Chap.~6]{ChaabanS2015}. Therefore, it is interesting to ask when non-adaptive transmission is optimal (in the capacity or the degrees-of-freedom sense) in multi-way settings.

\appendix
\section{Proof of Thm.~\ref{thm:gap_vanish}} \label{app:pf_of_med_gau}
Since $\Routt$ is rectangular and $\mathcal{C}$ is a convex region, it is sufficient to show that at least one point within an $O(1/\lambda_0^2)$ Euclidean ball centered at $(R_1, R_2) = (\Cowc(\sigma_1), \Cowc(\sigma_2))$ is achievable.

Let us consider the corner point $(\Cowc(\sigma_1), \Cowc(\sigma_2))$ in $\Routt$ and the corresponding one-way capacity-achieving distributions $p_i^*(x_i)$ with $\pi_i^* \triangleq p_i^*(1) = \min(\sigma_i, 1/2) > 0$. We will show that Shannon's inner bound at $p_1^*(x_1)p_2^*(x_2)$ achieves $(R_1, R_2) = (\Cowc(\sigma_1)-O(1/s^2), \Cowc(\sigma_2)-O(1/s^2))$. This implies that $\mathcal{C}(\sigma_1, \sigma_2)$ contains a point within distance $O(1/s^2) = O(1/\lambda_0^2)$ from the corner of $\Routt$.

Using the channel \eqref{eq:DMC_matrix}, Shannon's inner bound gives 
\begin{align*}
&\Delta \cdot R_1 \\
&\le I(X_1;Y_2|X_2) = H(Y_2|X_2) - H(Y_2|X_1, X_2) \\
&= p_2^*(0) H(Y_2|X_2=0) + p_2^*(1) H(Y_2|X_2=1) \\
&\qquad - \sum_{i, j \in \{0,1\} } p_1^*(i) p_2^*(j) H(Y_2|X_1=i,X_2=j) \\
&= (1-\pi_2^*) H(Y_2|X_2=0) + \pi_2^* H(Y_2|X_2=1) \\
&\qquad -(1-\pi_1^*)(1-\pi_2^*)H(Y_2|0,0) -(1-\pi_1^*)\pi_2^*H(Y_2|0,1) \\
&\qquad -\pi_1^*(1-\pi_2^*)H(Y_2|1,0) - \pi_1^*\pi_2^*H(Y_2|1,1) \\
&= (1-\pi_2^*) H_2( (1-\pi_1^*) As\Delta + \pi_1^*A(1+s)\Delta ) \\
&\quad + \pi_2^* H_2( (1-\pi_1^*) A(1+s)\Delta + \pi_1^*A(2+s)\Delta ) \\
&\quad -(1-\pi_1^*)(1-\pi_2^*)H_2(As\Delta) -(1-\pi_1^*)\pi_2^*H_2(A(1+s)\Delta) \\
&\quad -\pi_1^*(1-\pi_2^*)H_2(A(1+s)\Delta) - \pi_1^*\pi_2^*H_2(A(2+s)\Delta),
\end{align*}
where $H_2(p)$ is the binary entropy function, i.e., $H_2(p) \triangleq -p\log p -(1-p) \log (1-p)$.
Also note that
\begin{align*}
\Delta \cdot \Cowc(\sigma_1) &= H_2((1-\pi_1^*) As \Delta + \pi_1^* A(1+s)\Delta ) \\
&\qquad - (1-\pi_1^*) H_2(As\Delta) - \pi_1^* H_2(A(1+s)\Delta).
\end{align*}

Letting $\textsf{GAP}_1 \triangleq \Cowc(\sigma_1) - R_1$, we have
\begin{align*}
&\Delta \cdot \textsf{GAP}_1 \\
&= \pi_2^* H_2((1-\pi_1^*) As \Delta + \pi_1^* A(1+s)\Delta ) \\
&\qquad - \pi_2^* H_2( (1-\pi_1^*) A(1+s)\Delta + \pi_1^*A(2+s)\Delta ) \\
&\qquad -(1-\pi_1^*) \pi_2^* H_2(As\Delta) +(1-2\pi_1^*) \pi_2^* H_2(A(1+s)\Delta) \\
&\qquad + \pi_1^* \pi_2^* H_2(A(2+s)\Delta) \\
&= \pi_2^* H_2( A(s+\pi_1^*)\Delta ) - \pi_2^* H_2( A(1+s+\pi_1^*)\Delta ) \\
&\qquad -(1-\pi_1^*) \pi_2^* H_2(As\Delta) + (1-2\pi_1^*) \pi_2^* H_2(A(1+s)\Delta) \\
&\qquad + \pi_1^* \pi_2^* H_2(A(2+s)\Delta).
\end{align*}

Approximating by $H_2(p) \approx -p\log p + p$ for small $p$ and rearranging terms, we have
\begin{align*}
& \frac{\Delta}{\pi_2^*} \cdot \textsf{GAP}_1 \\
&= A \Delta \Big( -(s+\pi_1^*) \log(s+\pi_1^*) + (1+s+\pi_1^*) \log (1+s+\pi_1^*) \\
&\qquad + (1-\pi_1^*) s \log s - (1-2\pi_1^*) (1+s)\log(1+s) \\
&\qquad - \pi_1^*(2+s)\log(2+s) \Big) \triangleq A \Delta f(s).
\end{align*}

We will show that $s^2 f(s) \to \pi_1^*(1-\pi_1^*)/2$ when $s \to \infty$, which implies $f(s)$ decays as $\Theta(1/s^2)$.
\begin{align}
J &\triangleq \lim_{s\to \infty} \frac{f(s)}{1/s^2} \nonumber \\
&\stackrel{(a)}{=} \lim_{s\to \infty} \frac{f'(s)}{-2/s^3} = -\frac{1}{2} \lim_{s\to \infty} \frac{sf'(s)}{1/s^2} \nonumber \\
&= -\frac{1}{2} \lim_{s\to \infty} \frac{sf'(s)-f(s)+f(s)}{1/s^2} \nonumber \\
&= -\frac{J}{2} - \frac{1}{2} \lim_{s\to \infty} \frac{sf'(s)-f(s)}{1/s^2}, \label{eq:T_identity}
\end{align}
where (a) follows from l'Hospital's rule. Using $((s+c) \log (s+c))' = \log(s+c) + 1$ and rearranging terms,
\begin{align*}
&sf'(s) - f(s) \\
& = s \big( -\log(s+\pi_1^*) + \log(1+s+\pi_1^*) + (1-\pi_1^*) \log s \\
&\qquad + (1-2\pi_1^*) \log(1+s) -\pi_1^* \log(2+s) \big) - f(s) \\
&= \pi_1^* \log(s+\pi_1^*) - (1+\pi_1^*) \log(1+s+\pi_1^*) \\
&\qquad + (1-2\pi_1^*)\log(1+s) + 2\pi_1^*\log(2+s) \\
&\stackrel{(b)}{=} \pi_1^* \left( \log s + \frac{\pi_1^*}{s} - \frac{(\pi_1^*)^2}{2s^2} \right) \\
&\quad - (1+\pi_1^*) \left( \log s + \frac{1+\pi_1^*}{s} - \frac{(1+\pi_1^*)^2}{2s^2} \right) \\
&\quad + (1-2\pi_1^*)\left( \log s + \frac{1}{s} - \frac{1}{2s^2} \right) + 2\pi_1^*\left( \log s + \frac{2}{s} - \frac{4}{2s^2} \right) \\
&\quad + O(1/s^3) \\
&= \frac{3\pi_1^*(\pi_1^*-1)}{2s^2} + O(1/s^3),
\end{align*}

where (b) follows from the Taylor series of the natural logarithm at $s$:
\begin{align*}
\log(s+c_1) &= \log s + \frac{c_1}{s} - \frac{c_1^2}{2 s^2} + O(1/s^3).
\end{align*}

Hence we obtain
\begin{align*}
\lim_{s\to \infty} \frac{sf'(s)-f(s)}{1/s^2} = \frac{3\pi_1^*(\pi_1^*-1)}{2},
\end{align*}
and solving \eqref{eq:T_identity} gives
\begin{align*}
J = \frac{\pi_1^*(1-\pi_1^*)}{2}.
\end{align*}
Therefore we have
\begin{align*}
\textsf{GAP}_1 = \frac{A \pi_1^*(1-\pi_1^*) \pi_2^*}{2 s^2} + O(1/s^3).
\end{align*}

By symmetry, we also have
\begin{align*}
\textsf{GAP}_2 \triangleq \Cowc(\sigma_2) - R_2 = \frac{A \pi_2^*(1-\pi_2^*) \pi_1^*}{2 s^2} + O(1/s^3),
\end{align*}
which completes the proof.

\end{document}